\documentstyle[multicol,aps,prl,psfig,amssymb]{revtex}
\begin{document}
\title{Coulombic Phase Transitions in Dense Plasmas\footnote{Dedicated to the 70th birthday of Michael Fisher.}}
\author{W. Ebeling$^1$, G. Norman$^2$}
\address{$^1$Institut f\"ur Physik, Humboldt-Universit{\"a}t Berlin, Invalidenstrasse 110, D-10115 Berlin, Germany;\\
ebeling@physik.hu-berlin.de\\
$^2$ Institute for High Energy Densities, AIHT of RAS,
Izhorskaya street 13/19, Moscow 125412, Russia;\\
henry\_n@orc.ru\\}
\maketitle
\begin{abstract}
We give a survey on the predictions of Coulombic
phase transitions in dense plasmas (PPT) and derive several new
results on the properties of these transitions. In particular we
discuss several types of the critical point and the
spinodal curves of quantum Coulombic systems. We construct a
simple theoretical model which shows (in dependence on the parameter
values) either one alkali-type transition (Coulombic and van der
Waals forces determine the critical point) or one Coulombic
transition and another van der Waals transition. We investigate the
conditions to find separate Van der Waals and Coulomb transitions
in one system (typical for hydrogen and noble gas-type plasmas).
The separated Coulombic transitions which are strongly influenced
by quantum effects are the hypothetical PPT, they are in full analogy
to the known Coulombic transitions in classical ionic systems. Finally
we give a discussion of several numerical and experimental results 
referring to the PPT in high pressure plasmas.
\end{abstract}
\begin{multicols}{2}

\section{First order van der Waals and Coulomb transitions}

The theory of gases developed in the dissertation of van der Waals in 1873 may
be considered as the starting point of the modern theory of phase transitions.
Van der Waals´ approach is based on a simple physical model of interactions
between particles which takes into account short range repulsive as well as
long range attractive forces. The van der Waals model predicts below the critical
temperature the possibility of coexistence of two phases which differ from each
other by the density of molecules. In connection with the development of more
strict theories it became clear that van der Waals approach is restricted to
relatively weak attractive forces which either decay  faster than $1/r^3$ or
fulfil the Kac conditions. Therefore the applicability to Coulomb forces
remained open.

In 1943 Landau and Zeldovich discussed new possibilities
of phase transitions connected with metal-insulator transitions
in metal fluids in the vicinity of critical points \cite{ZL}. We
have to mention in this connection also the  work of Mott on
metal-insulator transitions in Coulomb systems. A first systematic study of
Coulombic transitions in plasmas was given  by Norman and Starostin
in 1968 \cite{NS68}. The first order transition predicted by these authors was
named plasma phase transition (PPT). The PPT was predicted as a possible
result from the competition between effective  Coulomb attraction and quantum
repulsion in the partially ionized dense plasma \cite{NS68,NS70}.  The
qualitative picture was similar to the van der Waals model where
the phase transition is a result from the long-range attraction between neutral
molecules and their short-range repulsion. We remember that the two phases in
the van der Waals model differ from each other by the  density of molecules.
The two phases in PPT have different number densities of
charged particles and different degrees of ionization. Atoms, which are present
in both coexistent phases, were treated originally as an ideal gas
\cite{NS68}.

To get the first estimate of the critical temperature Norman and
Starostin \cite{NS68} used the  thermodynamic functions available at that
time, namely  Debye-H\"uckel expressions for the chemical potential of the
charges with quantum corrections \cite{VL59}. The expressions described an
effective Coulomb attraction and an effective quantum repulsion due to the
uncertainty principle. In the linear approximation in the density a critical
temperature $T_{cr} = 2660$  and with the more realistic nonlinear
Debye-H\"uckel expression  the value $T_{cr}= 10640$ was obtained. We mention
that both expressions are valid only in the limit of small quantum repulsion
$(\lambda /r_D) << 1$  (here $\lambda$ is the electron thermal wavelength and
$r_D$ is the Debye radius), and  small non-ideality
parameter  $\gamma = e^2 n_i^{1/3} / kT << 1$,
where $n_i$ is the ion number density, $T$ is the temperature.
Later estimates of the PPT in hydrogen plasmas lead to higher values of the
critical temperature
[5-14]
 
The hypothetical phase transitions in multiple ionized plasmas were
treated first in \cite{VMN}. A detailed study of He-plasmas
 was given in
\cite{FoKaEb}. Here we will consider only single-charged
 ions in gaseous
plasmas.
 
Since up to now the PPT in plasmas is not yet clearly identified
experimentally we plan to make as clear as possible the definition,
the properties and the conditions for a PPT. For this reason we
develop a simple model (a combination of the Debye-H\"uckel with
the Van der Waals theory) which shows a separate Coulombic
transition - which is a PPT. Other related
phenomena in plasma systems are discussed only in brief.

Plasma-like phase transitions in the optically excited electron-hole
system in semiconductors were studied with similar methods
\cite{Ebel71,IN,KEK75,EKK76,leeb96}. The experimental observations
seem to point out that these instabilities of the theory correspond
to a real phase transition in semiconductors, this topic will not
be considered here in detail.

In more detail we will discuss comparison of the PPT to the known
transitions in classical Coulombic systems. The first results on phase
transitions in ionic systems go back to the late sixties.
In 1970 Voronzov et al. \cite{Voronzov} found a coexistence line
and a critical point in the course of Monte Carlo studies of charged
hard spheres imbedded into a dielectric medium. An analytical estimate
of the critical point and the coexistence line for electrolytes based
on the Debye-H\"uckel theory was given in a short note by one of the
authors \cite{Ebe71}. However a systematic theory of classical
Coulombic phase transitions including a comparison with numerical and
experimental data was given only in the pioneering work of
Michael Fisher and his coworkers \cite{levin_fisher}. These
authors have also made an extensive investigation of the state of art
in this field \cite{Fish}.
Summarizing these results for classical systems we may say that theory and
experiment (Monte Carlo data as well as measurements on electrolytes) are in
quite good agreement. There is no doubt any more that a Coulombic phase
transition in classical systems exists. We will not repeat here all the
arguments but refer to the careful investigations of  Fisher et
al. \cite{levin_fisher,Fish}.

The classical Coulombic transition is due to a balance between
a hard-core repulsion and a Coulombic attraction.
We will show here that the PPT is a balance between quantum
repulsion between point charges and Coulombic attractions.
In this respect PPT is a kind of quantum variant
of the classical Coulombic transition.
If on one hand the existence of a Coulombic transition
in ionic systems is now well confirmed, there is on the other hand
no proof yet for the existence of a PPT in plasmas. At present the
problem of the existence of a PPT is still open.
From the point of view of the theory this is connected
with the difficulties to derive an accurate equation of state
for nonideal quantum plasmas. From the experimental point of view
the difficulties are connected with the very high pressures where the
PPT could occur. However as we will point out here there are now
several experimental and numerical data
which point to the existence of a PPT in real plasmas.

\section{Debye H\"uckel approximation for classical and quantum
systems}

In order to describe Coulombic phase transitions one needs good approximations for the
thermodynamic functions in the regions where - according to the first estimates - the phase
transition is to be expected. As we have learned from the theory of phase transitions in neutral
gases, already simple expressions as the van der Waals equation might give qualitatively correct
results. We will show here that the Debye-H\"uckel expressions for the
thermodynamic functions together with the mass action law provide an
approximation which might serve a a zeroth approximation for the description
of Coulombic phase transitions. In their first work Norman and Starostin
\cite{NS68} proposed for a first estimate of the critical temperature, to use
Debye-H\"uckel-type expressions $(1 - c \lambda /r_D)$ for the chemical
potential. Here $\lambda$ is electron thermal wavelength, $r_D$ is a Debye radius,
$c˜\simeq 0.1$ is a numerical coefficient. The Debye-H\"uckel term
describes an effective Coulomb attraction, the $(\lambda/r_D)$-term represents
in this approach an effective quantum repulsion due to the uncertainty
principle.  Norman and Starostin \cite{NS70} used also the original expression
of Debye- H\"uckel for the chemical potential $(1 + c\lambda /r_D)^{-1}$
which has  the same accuracy if $(\lambda/r_D)$ tends to zero.

We will give now a systematic discussion of the Debye-H\"uckel
approximation for the classical and quantum cases. We consider a binary
Coulomb system with $n_+$ positive ions (cations) and $n_-$ negative charges
(anions or electrons) per cubic centimeter $n_+ = n_-$. The density of neutrals
is $n_0$. The total density is $n = n_+ + n_0 = n_- + n_0$. In the
following we will use the plasma notations, i.e. we call the negative charges
"electrons" and the positive charges simply "ions". In
the Debye-H\"uckel theory for charged hard spheres with the diameter $a$ the
excess chemical potential of the charges of species $i$  is given by the
simple expression
\begin{equation}
\mu_i^{ex} = -\frac{e_i^2}{D_0 kT (r_D + a)}
\end{equation}
The densities of free and bound particles are connected by the
mass action law (Saha equation)
\begin{equation}
\frac{n_0}{n_i n_e} = K(T) \exp\left(  \frac{\mu_{ex}}{k_B T} \right)
\label{saha}
\end{equation}
where $K(T)$ is the mass action constant which is given by Bjerrum's
formula or certain modifications \cite{Ebel71,EbHiKr02} in the classical case.
The osmotic pressure is given by the formula
\begin{equation}
P / kT  = n_i + n_e + n_0 - \frac{1}{24 \pi} \kappa^3 \phi(\kappa a)
\end{equation}
Here the Debye-H\"uckel function $\phi(x)$ is defined by
\begin{equation}
\phi(x) = \frac{3}{x^3}\left[1 + x - \frac{1}{1+x} - 2 \log(1+x)\right]
\end{equation}
The easiest way to transfer the Debye-H\"uckel approximation to quantum plasmas
is the replacement of the hard sphere diameter by an effective
quantum diameter of charges. According to the Heisenberg principle
a point particle with the thermal momentum
\begin{equation}
\hat p = \sqrt{\frac{8 m kT}{\pi}}
\end{equation}
corresponds to a wave packet of size
\begin{equation}
\delta x = \frac{\hbar}{\hat p} = \frac{\pi \hbar}{4\sqrt{2\pi mkT}}.
\end{equation}
Identifying this with the effective diameter of the charges we get
\begin{equation}
a \rightarrow a(T)  = \frac{\Lambda}{8} = \frac{h}{8 \sqrt{2\pi m_e kT}}
\end{equation}
In this so-called Lambda approximation
\cite{EKK,EbFo91} quantum effects are expressed
by just one characteristic length, the thermal de Broglie wavelength
$\Lambda$ in combination with Debye-H\"uckel type approximations.
The Lambda approximation agrees very good with the
exact quantum-statistical results for small densities for $T < 10^5 K$ \cite{EKK,KKER}.
This approximation requires that the plasma is nondegenerate
i.e. $ n^*\Lambda^3<<1$ (where $n^*$ denotes the density of free particles).
Furthermore only a small fraction of the atoms should be
bound in molecules $\beta_2 << 0.5 $. The standard choice of the
mass action constant for quantum plasmas is the Brillouin-Planck-Larkin
expression
\begin{equation}
K(T) = \Lambda^3 \sigma(T)
\label{K}
\end{equation}
where
\begin{equation}
\sigma(T) = \sum_{s=1}^{\infty} s^2 \left(\exp(-\beta E_s) -1 + \beta
E_s\right)
\end{equation}
In this way the classical and the quantum case might be
treated in the present rough approximation by the same procedure.
The easiest way to check for stability of the system is the investigation of
the sign of the derivative of the pressure. The region where
\begin{equation}
\frac{\partial \beta P}{\partial n} < 0
\end{equation}
corresponds to thermodynamic instability. Equivalent is the
condition \cite{NS68,NS70}
\begin{equation}
\left({\frac{\partial n_i}{\partial n}}\right)_{T}  < 0
\end{equation}
which is valid only if atoms are treated as ideal gas.
Instability is restricted to temperatures below the critical one which is
defined by
\begin{equation}
T \le T_{cr} = 16 \frac{e^2}{D_0 k_B a}
\end{equation}
In the case of classical ionic solutions the parameter $a$ is to be
identified with the diameter of ions. In the case of dense quantum plasmas 
we have to introduce the quantum effective diameter $a(T) = \Lambda/8$. 
This way we have shown that the PPT is the analogue of the known
classical Coulombic transition in ionic solutions.
At least in the simplified theory given above the classical transition
in ionic solutions and the quantum PPT in dense plasmas are in 
full analogy. 
In both cases the spinodal curve is
given by
\begin{equation}
\mu_{1,2} = \left(\frac{b^2}{8}-b\right) \pm
\left[\left(\frac{b^2}{8}-b\right)^2 - b^2\right]^{1/2}
\end{equation}
where $b = e^2 / k_B T a(T)$ and $\mu = e^2 \kappa / k_B T$.
In the quantum case we find
the corresponding estimate for the critical temperature
\begin{equation}
T_{cr} \simeq  12570 K
\end{equation}
In the case of electrolytic solutions or semiconductors the critical
temperature is much lower due to the appearance of the dielectric constant in
the denominator. Another effect which decreases $T_{cr}$ is the finite size
of ions. In order to describe the size of the ion cores in
alkali or noble gas plasmas we introduce an effective ion radius $R$,
e.g. we have $R \simeq 1.69 A°$ for $Cs-$ ions. Further we replace
$\Lambda/8$ by
\begin{equation}
a(T) = \Lambda/8 + R
\end{equation}
Assuming that $R$ is temperature-independent we find now from Eq.(15)
\begin{equation}
T_{cr} =  12570 K \frac{2}{1 + \left(1 + (6 R/\pi a_B) \right)^{1/2}}
\end{equation}
where $a_B = \hbar / e^2 m_e$ is the Bohr radius. We see that a finite ion size
may lead to an essential decrease of the critical temperature to values
far below $10000 K$. For $Cs-$ plasmas this estimate gives e.g.
$T_{cr} \simeq 6000 K$. Since this value is still far above the
observed critical point $T_{cr} \simeq 2000 K$ we have to search now for
additional effects, which might reduce the critical temperature.

\section{Combination of van der Waals and Debye-H\"uckel approximation}

In the Debye-H\"uckel type approximation derived in the previous section the
interaction of charges with neutrals and the neutral-neutral interaction was
not taken  into account. This model might
fail if the interactions with neutrals have
an essential influence. In the following we consider for concreteness
a plasma system. We treated the neutrals as ideal particles in
the simple approach given above which is analytically solvable. The
system of mass action laws might become rather difficult if one includes the
interactions with neutrals or/and chemical equilibria
between neutrals as e.g. between atoms and molecules.
A better starting point for such more complicated cases
is as a rule the expression for the free energy, including a
minimization procedure.

Combining the Debye-H\"uckel model with a van der Waals expression
and a contribution which takes into
account the polarization of neutrals we get for the free energy
\begin{eqnarray}
  \label{fenergy}
F&=& n_0 k_B T \left[ \ln\left(\frac{n_0 \Lambda^3_0}{\sigma(T)}\right) - 1
\right]
 + n_i k_B T \left[ \ln\left(\frac{n_i \Lambda^3_i}{g_i} \right) - 1 \right]
 \nonumber \\
&+& n_e k_B T \left[ \ln\left(\frac{n_e \Lambda^3_e}{g_e} \right) - 1 \right]
 - k_B T V \frac{\kappa^3}{12 \pi} \,
  \tau\left( \frac{\kappa \Lambda}{8} \right)
\nonumber \\
 &-& k_B T n_0 \ln ( 1 - n_0 B) - A n_0^2  - W n n_0
\end{eqnarray}
Here the last three terms denote the
contributions from the  short range van der Waals interactions  and the
contributions from  polarization terms. The polarization
terms are of particular
importance for the description of alkali and mercury plasmas
\cite{hensel,redmer}.

Three free constants characterize now our model\\
- the repulsion of atoms : $B$          \\
- the attraction of atoms : $A$ \\
- the strength of atomic polarizability : $W$

In order to reduce the number of free parameters we assume in the
following $W=0$. We note that the free energy in this form defines
a model plasma with properties intermediate between van der Waals
gas and a Debye-H\"uckel plasma.
The equilibrium composition is given by the nonideal Saha equation
\begin{equation}
\frac{1-\alpha}{\alpha^2} =
n \Lambda^3 \sigma(T) \exp\left( - \frac{\Delta I(\alpha)}{k_B T} \right)
\label{saha1}
\end{equation}
Here
\begin{eqnarray}
  \label{lowering}
\Delta I(\alpha) &=&
\frac{e^2 \kappa \sqrt{\alpha}}{k_B T (1+\kappa a(T) \sqrt{\alpha})}
- \frac{(1-\alpha)n B}{1-(1-\alpha)n B}\nonumber\\ 
&+& \log(1 - (1-\alpha) n B)
\nonumber \\[.2cm]
&+& (1 - \alpha) n A - W n (2 \alpha - 1)
\end{eqnarray}
is the lowering of the ionization energy.

As well known, real gases with attractive interactions show a first-order
phase transition described in the $p-T$ plane by a critical point $C_1$ and
a coexistence line ending in $C_1$. Assuming that the Coulombic terms are
omitted in our expressions for the free energy
a simple van der Waals type expression is obtained with the
critical point
\begin{equation}\label{tc1}
T_{vdW} = \frac{8 a}{27 k_B b}, n_{vdW} = \frac{1}{3 b}
\end{equation}
The critical temperature is about $10^1 < T_{cr} < 10^3$
for realistic values of the parameters. For example $T_{cr} \simeq 33 K$
for hydrogen. The degree of ionization is
practically zero in this region, i.e. there is no interference with the PPT.
On the other hand, at least for hydrogen, the PPT occurs in a region
where the number density of neutrals is
rather low. We may assume therefore that the assumption
$a = 0, b =0$ is justified in this case. Then, as shown above, the PPT may be
treated in an analytical way.
In the Lambda-approximation we got in the previous section
an analytical expressions for the critical point of the PPT
\cite{EKK,leeb96}. By
introducing the Bohr radius $a_B$ we may transform the
expression for the critical temperature  to the form
\begin{equation}\label{tc2}
T_{ppt} = \frac{e^2}{8 k_B a_B}
\end{equation}
Further the critical density of the free electrons may be written as
\begin{equation}
n_{ppt} = a_B ^{-3}
\end{equation}
What happens in a system where both types of interactions
are present? Evidently now besides
the classical first-order phase transition typical for neutral gases
a second first-order phase transitions due to Coulomb forces
may appear. The second one which
might occur only at rather high temperatures $T > 10^4 K$ is the PPT.
In Fig. 1 we give the pressure isotherm of hydrogen plasmas at $T = 10000 K$.
\begin{minipage}{8cm}
\begin{figure}[htb]
  \psfig{file=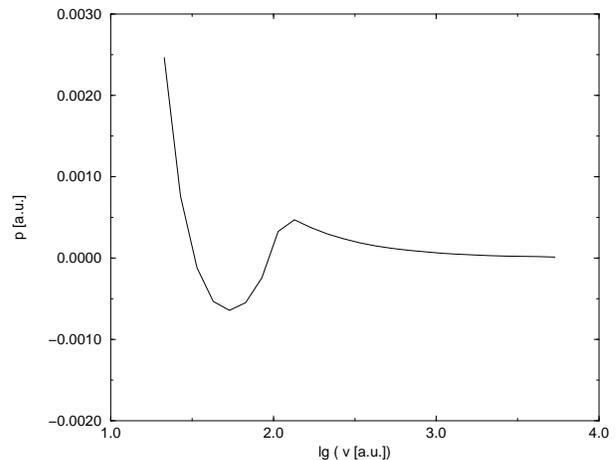,angle=-90,width=8cm}
  \caption{The isotherm $T = 10000 K$ for hydrogen plasmas showing the PPT instability.}
  \label{fig:skizze}
\end{figure}
\end{minipage}

In the unstable region of the isotherm the plasma is separated into
two phases. The dense phase is highly ionized and the Coulomb interaction
dominates over the van der Waals forces. The less dense phase is only
weakly ionized and van der Waals interactions dominate. A more realistic
calculation of the coexistence line for hydrogen plasmas was given recently
\cite{BEF99}.  A coexistence pressure of $p \simeq 115 GPa$ was found in
the cited work for $T = 10000 K$; the corresponding mass densities of the
coexisting phases are $\rho_1 \simeq 0.62 g/cm^3$ and
$\rho_2 \simeq 0.82 g/cm^3$.

In order to
investigate the qualitative influence of the van der Waals forces we studied a
model plasma with increasing values of the parameters $a$ and $b$. Fig. 2
shows the isotherm $T = 0.07$ a.u. and $a = b = 30$ a.u.. 
\begin{minipage}{8cm}
\begin{figure}[htb]
\psfig{file=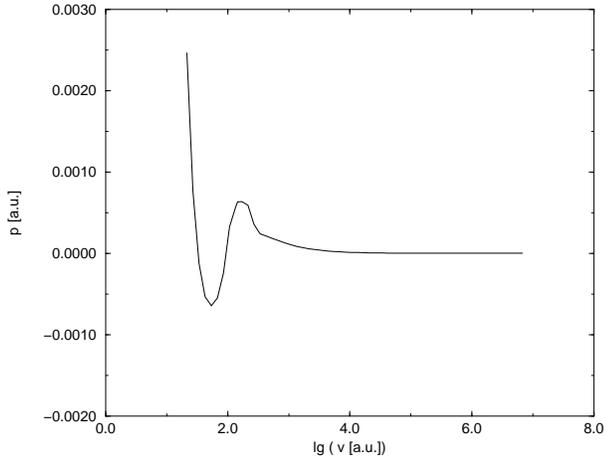,angle=-90,width=8cm}
  \caption{The isotherm $T = 0.07$ a.u. for an alkali-like
model plasma with one transition of mixed
PPT-van der Waals type (van der Waals parametersof the model:
$a = 30$ a.u., $b = 30$ a.u.).}
  \label{2}
\end{figure}
\end{minipage}

In spite of the
fact that these values are rather large we cannot detect a separate van der
Waals wiggle. Both attracting forces support each other and lead to one phase
transition approximately in the same density region as the PPT. This is the
situation we observe in nature for alkali plasmas. At further increase of the
values of the van der Waals parameters to $a = b = 100$ a.u. we observe two
separate van der Waals loops as shown in Fig. 3. This is the situation we
observe for real noble gas plasmas. We are to emphasize, that Fig. 2 and Fig. 3 have
just illustrative character; in reality we do not see two wiggles in one
isotherm due to the drastic difference between the critical temperatures.

Summarizing these findings we may state that in dependence on the values
of the van der Waals parameters we may obtain either
a phase diagram with two first order transitions (hydrogen and noble gas
plasmas) or a phase diagram where both transitions fuse to just one
(alkali plasmas). The
existence of a phase diagram including a van der Waals type transition  and a
separate metal-insulator phase transition was discussed for the first  time in
1944 by Landau and Zeldovich \cite{ZL}.
We repeat that the existence of two separate phase transitions requires
the validity of the inequalities
\begin{equation}
T_{vdW} << T_{ppt};    \rho_{vdW} << \rho_{ppt}
\end{equation}
This is fulfilled for hydrogen and for noble gas plasmas. In the case
of alkali plasmas the van der Waals and the PPT transition
have about the same critical density and temperature. Therefore
both transitions fuse just to one with very specific properties
\cite{KN00,hensel,redmer,warren}.
\begin{minipage}{8cm}
\begin{figure}[htb]
\psfig{file=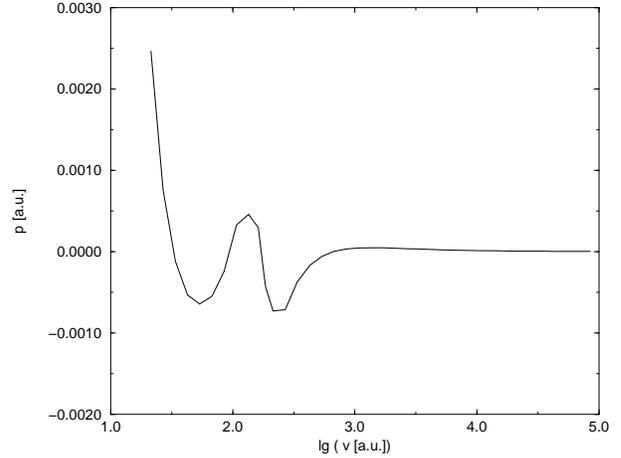,angle=-90,width=8cm}
  \caption{The isotherm $T = 0.07$ a.u. for a noble gas-like
model plasma with a PPT and a separate van der Waals
transition (van der Waals parameters 
of the model: $a = 100$ a.u., $b = 100$ a.u.).}
  \label{3}
\end{figure}
\end{minipage}

\section{Discussion of Plasma Phase Transitions}

Considerable efforts were applied to obtain better plasma thermodynamic
functions for larger
range of the nonideality parameter. We will not go into the details of the
theory here which are explained elsewhere \cite{EKK,KKER,EbFo91,SC89}.
The critical temperatures
derived from more refined versions of the theory lay in the region
$T_{cr} = 14000-19000 K$ \cite{EKK,ER85,SC89}. There is  a scatter in
the values of $T_{cr}$, $p_{cr}$ and $n_{cr}$, since the
procedure of getting the thermodynamic functions for large nonideality is not
a unique one.  Instead of going into details on the several theoretical
approaches we concentrate here on the physics and the relation to experiments,
to numerical simulations and to related phenomena.

Let us start with the experimental aspects. The existence of a PPT
in dense plasmas is still less clear than in classical
Coulombic systems. This is  mainly due to the fact that
pressures predicted for the coexistence line are above $1 Mbar$ what is  still
hardly reachable in experimental situations. However in the last time more
and more experiments were performed which cover the region of interest
\cite{Weir,Fortov,Nellis,Silva,Collins,FT,Caub,Most}.
Experiments with shock-compressed hydrogen and deuterium plasmas
have verified that around $140 GPa$ and $3000 K$ a transition to a highly
conducting state occurs \cite{Weir,Nellis}. This is approximately the pressure
region where the coexistence line of the PPT is expected \cite{BEF99}.
A corresponding behavior of conductivity data has been
reported \cite{Fortov}.
Pressure dissociation  and ionization, which almost do not depend on the temperature,
become a dominant factor at $10^4 K$ and below in hydrogen and deuterium fluid
studied experimentally at high pressures \cite{Nellis,FT,Caub,Most}.

In order to compare theory and experiment in a more quantitative way
refined treatments of the  phase transition induced by
pressure dissociation  and ionization were given
\cite{EKK,ER85,SC89,SC00,KN00,BEF99}. Most of the possible
species and various
interactions were taken into account to calculate the
equations of state. There is a certain
discussion whether the model
equations of state reproduce reasonably well the recent experiments. However
the theory  predicts in all variants a phase transition.
with $T_{cr} = 14000-19000 K$. Excited states do not
contribute significantly to the hydrogen partition function in the
intermediate range of temperatures.

Path integral Monte Carlo calculations and wave packet dynamics
also give some hints to the existence of the plasma phase
transition \cite{Zamalin,KTR94,ceperley,ceperley96,mil-pol,mil-cep,fiboebfo01}.
For example the validity of the thermodynamic expressions used was tested by
comparison with molecular simulation and experimental data for nonideal
plasmas and limiting expansions for weakly nonideal plasmas
\cite{fiboebfo01}. The problem continues to attract remarkable attention,
for example at the recent (June 2-7, 2002) ``International Conference on
Warm Dense Matter and FEL Experiments'' in Hamburg,
Filinov et al \cite{FFBL,FBF} claim that they ``present results of direct
path integral Monte Carlo simulations which, for the first time, provide
first-principle support of such a phase transition in dense hydrogen''.
It should be noted, however, that these particular PIMC simulations 
yield rather low energy values for the ``droplets'' that appear in the
simulations. Kohanoff and Scandolo \cite{KoSc} discussed on this workshop 
the same problem in a softer manner and presented their results of ab 
initio molecular dynamics simulations. We mention these works as part of the
ongoing effort to simulate the PPT.

Let us discuss now in brief the phase transitions observed in
alkali-plasmas. As pointed  out in previous sections the theory predicts for
this case that the PPT and the van der Waals transition fuse just to one
transition. Near to the critical point the degree of ionization is
low but different from zero. Coulombic interactions are present
and influence the critical behavior, in spite of the fact that
they do not play the dominant role. The area of parameters close
to the critical point of cesium was carefully investigated \cite{hensel}.
It was shown that near to the critical point the ionization landscape
changes quickly and has a quite complicated structure.
The theoretical predictions which take into account
the charge-atom interaction (polarization effects) are in rather
good agreement with the data \cite{redmer,KN00}.

The existing experiments for argon
plasmas based on shock wave data are also in satisfactory
agreement with the theory \cite{FT,FL1,FL2}.

We will discuss now the problem of critical behavior and the critical
indices. For the classical case much attention has been devoted to this problem
\cite{Fish,Fish2,Br}. For the PPT where quantum effects influence the critical
point, the problem of critical indices is still open. So far no experiments
are available which give reliable information about the critical
properties of hydrogen and noble gas plasmas.
Experimental studies of classical electrolyte solutions revealed that Coulomb
liquids  characteristics differ from those of simple liquids. Even if
crossover was observed from classical scaling laws to Ising scaling, it takes
place much closer to the critical point. The Ginzburg parameter, which
characterizes the size of the Ising region, turns out to be of 1-2 and more
orders of magnitude less than for simple liquids. So there is a remarkable
difference in crossover range between simple and Coulomb
liquids \cite{Fish,Fish2,Br}. Experimental study of crossover phenomena in real
plasmas and in particular in cesium plasmas
might help to find out if the critical point is of the plasma or gas-liquid
kind. Additional factors, which do not exist in electrolytes, namely,
quantum effects and charge-atom interaction can influence the crossover phenomena
in PPT.

Michael Fisher \cite{Fish2} formulated clearly the challenge to the
critical phenomena theory:  "Critical behavior is thoroughly understood in
Ising models alias lattice gases: But how far does that help our insight into
the critical region of continuum models and real fluids? In particular, what
causes some ionic solutions to exhibit van der Waals or "classical" critical
exponents, while other, so-called "solvophobic" systems are of Ising - type?".
 It is evident that the difference between properties of various liquids under
similar external conditions can be attributed principally only to the
interaction potentials between the particles. However the
Kadanoff-Wilson theory of critical phenomena, which is based on the
universality hypothesis, neglects interaction potential peculiarities and
appeals only to large-scale properties, such as system dimension and
Hamiltonian symmetry. So other approaches are needed.

First of all Fisher points to the
integral equation approach. But most of the standard approaches to the theory
of critical phenomena based on integral equation method fail \cite{Fish2}.
Though both the Kadanoff-Wilson theory and the integral equation approach are
on equal rights, the first is based on global treatment whereas the
latter is related
to the local Ornstein-Zernike equation. For this reason Fisher
puts the integral equation approach on the first place in his hope
to solve the
problem of Coulomb criticality. However there is a specific difficulty in
the local approach which is connected with the choice of closing
relation in the integral equation approach. This difficulty was attacked
recently by Martynov \cite{Mart1}. He suggested an integral equation which
depends on the interaction potential and permits in principle to define the
critical indices and their dependence on the interaction potential. Martynov
\cite{Mart1} applied his local approach to the system of charged particles.
The critical indices obtained differ almost as much as twice from van der
Waals and Kadanoff-Wilson predictions. However the results cannot be
applied directly to electrolyte solutions since they do not take into
account solvent molecule influence, neither it can be applied directly
to dense plasma, since quantum effects are not taken into account.

Let us finally discuss several special problems connected with
Coulombic phase transitions. At first we mention the
magnetic field influence. It was shown recently that
strong magnetic fields increase the critical temperature \cite{ESO,PCS}.
Another important effect is connected with nonequilibrium phenomena.
It was noted \cite{NV} that nonideal plasmas are usually generated in
non-equilibrium state with  respect to plasma wave excitation. This might
complicate the theory \cite{KN00} and the analysis of  experiments.

At third we would like to discuss a new class of phenomena
observed in Cs-plasmas \cite{Holmlid}. Though the standard cesium phase
diagram is well known \cite{warren}, Recently a new phase state of Cs was
observed by Holmlid et al \cite{Holmlid,MOP01,MOP02}.
In experiments with thermoionic converters
clusters were observed first at $1300 K$. These cluster were cooled
and associated in microdroplets. The
microdroplets formed were fixed and measured at $70 K$. The number
density obtained was about $1018$ $atoms/cm^3$.
Holmlid \cite{Holmlid} considered his results as an observation of
the Rydberg  matter predicted by Manykin et al \cite{MOP01,MOP02}.
Generally speaking Rydberg matter exists only at $0 K$ temperature.
At higher temperatures Manykin's approach can and should be combined with our
non-ideal plasma treatment. In fact both approaches enhance each other.
The phenomenon might be connected with the idea of isolated segment of
metastable nonideal plasmas which was introduced into the
plasma thermodynamics long ago \cite{N71}. Supercooled nonideal plasma
states were a direct consequence of the PPT \cite{N00}.  Manykin's Rydberg
matter \cite{MOP01,MOP02}, Holmlid's microdroplets \cite{Holmlid} and
plasma phase transitions \cite{NS68,NS70,N01} are various facets of
Coulombic phenomena. The existence of isolated regions of
metastable nonideal plasmas is not an exceptional feature of cesium.
Several authors \cite{N71,N00,MOP01,MOP02} suggested that this might be
a more general  feature. In fact Holmlid \cite{Holmlid} observed his
clusters in many other substances as well. The isolated region of metastable
nonideal plasmas is possibly an inherent part of the phase diagram.
This part complements the standard diagram and is superimposed on it.

\section{Conclusions}

This work is devoted to plasma phase transitions (PPT)
which are determined by attractive Coulombic interactions and
quantum repulsion. The PPT occurs in the region
of high pressures and temperatures, where matter is at least
partially ionized. At least in principal the PPT is an analogue of the known
Coulombic transition in classical ionic solutions. In both cases the
mechanisms are the same, the transition is due to the interference
between a Coulombic attraction of the charges and repulsive forces. The
classical repulsion of ions at small distances is in the quantum case
replaced by Heisenberg and Pauli repulsion effects. We have shown above
for dense gaseous plasmas that in dependence on the values of the van der Waals
parameters of the neutral component we may obtain either a phase diagram wi
two first order transitions (hydrogen and noble gas plasmas) or a phase
diagram where both transitions fuse to just one (alkali plasmas). In spite of
considerable efforts there is no final proof yet that the PPT really exists,
however new theoretical and experimental results seem to confirm this
hypothesis. One strong theoretical argument in favor of the PPT is the
analogy to the Coulombic phase transition in classical systems. For ionic
solutions theory and experiment (Monte Carlo data as well as measurements on
electrolytes) show a corresponding Coulomb transition. Therefore there is no
doubt any more that a Coulombic phase  transition in classical systems exists
\cite{levin_fisher,Fish,Fish2}. For quantum systems as e.g. hydrogen plasmas
the problem of the existence of a PPT is still open. The main reason for
still unsolved problems are 
\begin{itemize} 
\item the theory of the PPT
requires an accurate theory of dense quantum plasmas, and 
\item the pressures
where the PPT occurs are in the Mbar region what leads to very difficult
experimental problems. 
\end{itemize}
We discussed here several results of theory and
experiment which point to the
existence of a PPT in real plasmas and discussed the conditions and parameter
regions where the PPT is expected.

\section*{Acknowledgements}
We thank V.E. Fortov, A.S. Kaklyugin, E.A.Manykin and two referees
for valuable discussions.
The work is supported by RFBS, grant 00-02-16310a.

\end{multicols}
\end{document}